\documentclass[12pt]{article}
\usepackage{graphicx,latexsym}
\newcommand{\bo}{\raise-1mm\hbox{\Large$\Box$}}   
\begin{document}

\begin{titlepage}

\begin{center}
{\Large \bf Center vortices, the functional Schr\"odinger equation, and
CSB}\\[.2in]
Talk at the symposium {\em Approaches to Quantum Chromodynamics},
Oberw\"olz, September 2008\\[.2in]
John M. Cornwall\footnote{Email:  Cornwall@physics.ucla.edu}\\
Department of Physics and Astronomy,\\
 University of California, Los Angeles CA 90095

\end{center}

\begin{abstract}
 The functional Schr\"odinger equation (FSE) for QCD gives a unique
perspective on generation of a gluon mass $m$, as required for center
vortices.  The FSE, which yields a special d=3 gauge action, combined with
lattice calculations strictly in d=3 give a value for the dimensionless
ratio of d=3 coupling to mass $g_3^2/m$. From this we infer a reasonably
accurate value for the d=4 running coupling $g^2(0)$ in the region of low
momentum where it is nearly constant.   The result, consistent with other
estimates, is too low to drive chiral symmetry breaking (CSB) for quarks in
a standard gap equation that has no explicit confinement effects.  We
recall and improve on old work showing that confinement implies CSB for
quarks, and consider CSB for test (that is, quenched) Dirac fermions in the
adjoint representation.  Here the previously-found value of $g^2(0)$ is
large enough to drive CSB in a gap equation, which we relate to the
presence of center vortices (non-confining, for the adjoint) and nexuses
that drive fermionic zero modes.  We discuss the extension of adjoint CSB
to finite temperature.
\end{abstract}

\end{titlepage}

\section{Introduction}

I discuss two related topics. The first is an approximate evaluation of the
QCD coupling at zero momentum,  $\alpha_s(0)$, using the functional
Schr\"odinger equation (FSE).  The second is the beginning of a program,
not completed, for relating chiral symmetry breakdown (CSB) properties for
both quarks and (hypothetical) adjoint Dirac fermions in QCD to
$\alpha_s(0)$.  This program can be looked on as using known CSB  results
(from the lattice) to constrain the allowed range of $\alpha_s(0)$ or as
using a theoretically-determined value of $\alpha_s(0)$ to predict the CSB
properties.  The present accuracy of some preliminary investigations of
these issues is not high enough to be definitive, but it does suggest that
$\alpha_s(0)$ in the range 0.6$\pm$0.2 is consistent with known
properties of CSB.  

 The FSE has long been of interest
to me \cite{corn88}; more recently,  motivated by a talk \cite{olejnik} of
\u{S}.
Olen\'{\i}k at the 2006 Oberw\"olz symposium, I wrote a paper last year
\cite{corn07}, one conclusion  of which is an estimate of the
$d=4$ strong coupling constant\footnote{Please note that this coupling is
the  scheme- and
process-independent coupling defined in the pinch technique
\cite{corn76,corn81,corn82,cornpap,binpap,abp,abp1} and not the
process-dependent
running coupling that is related to our coupling by a process-dependent
transformation \cite{brodsky}.  The pinch technique is an all-order way of
extracting from Feynmans graphs for the S-matrix new graphical structures
for off-shell proper vertices that are completely locally gauge-invariant.}
   at zero (or small) momentum,
$\alpha_s(0)\equiv g^2(0)/(4\pi )\simeq 0.5$. This estimate is somewhat
unusual since this
estimate is based on one-loop gluon gap equations and lattice numerics in
$d=3$, not $d=4$, plus some theory
that attempts to relate $d=3$ QCD to $d=4$ QCD through the FSE.  The $d=3$
studies  
\cite{chk,cornyan,alexn,buchphil,corn1998,eber,karsch,hkr,ckp,naka}
vary somewhat, but in my view (Ref. \cite{corn1998} argues that the one-loop
gap-equation results for $m$ are too small) the most reliable value for the
ratio $g_3^2/m$ is $\simeq 6.3/N$ (for gauge group $SU(N)$), where
$m$ is the dynamical gluon mass and $g_3^2$ is the $d=3$ gauge coupling. 
FSE theoretical estimates suggest $g^2(0)/(4\pi )\simeq (2.9/4\pi
)(g_3^2/m)\simeq 0.5$.
There are, of course, other estimates of $\alpha_s(0)$:   Phenomenology
sensitive to
infrared properties of QCD gives $\alpha_s(0)\simeq 0.7\pm 0.3$
\cite{natale}.  Other  pinch-technique calculations \cite{abp,abp1} suggest
$\alpha_s(0)\simeq 0.5$, just as I use here.

Recently, lattice simulations \cite{perez} have been reported for an
interesting question with a long lineage \cite{engels,basile,karsch2}:  How
does CSB work in QCD with adjoint fermions\footnote{The lattice work is, of
course, in Euclidean space, where there are no Majorana fermions and hence
adjoint fermions plus gauge fields are not a supersymmetric theory.}
instead of the usual fundamental-representation quarks?  These authors'
works state that there is CSB for quenched adjoint fermions not only at
temperatures below the deconfinement phase transition but also up to a
temperature exceeding the deconfinement phase transition.  In contrast, for
quarks there is no CSB above the deconfinement phase transition, consistent
with lattice findings \cite{deforc} that center vortices, the standard
confinement machinery of today, are   both necessary and sufficient for CSB
with quarks.   Adjoint fermions, blind to the long-range parts of center
vortices, are not confined and the
argument is, therefore, that there are different mechanisms for CSB for
fundamental and adjoint fermions.  I do not claim to have definitively
answered this question, but present numerical estimates based both on
estimates of $\alpha_s(0)$ and on fermion gap equations that have
appropriate kernels for small momentum suggest an answer that, given fairly
large uncertainties, seems to accord with present-day knowledge from the
lattice.  What I report here is at best the beginning of a program of
refining our quantitative understanding of QCD by comparing
 estimates of $\alpha_s(0)$ with lattice and theoretical studies of
adjoint CSB.   

  Theoretical papers from long ago  \cite{corn80,johnson,casher}
argue that various confinement mechanisms produce CSB for quarks.  (Later I
will give a brief update of some of these ideas, based on the role of a
condensate of center vortices and their close relatives, nexuses, in
confinement and CSB).  These confining mechanisms depend only indirectly on
$\alpha_s(0)$.  In principle it could be (although the lattice data say
otherwise) that there would be CSB for quarks even with no confinement, a
possibility that does depend on $\alpha_s(0)$.  Generally, CSB for quarks
and adjoint fermions should be sensitive to three    couplings:  (1) the
standard QCD coupling $g^2(0)/(4\pi )$ at zero momentum transfer; (2) the
critical coupling $g_c^2(0; fund)/(4\pi )$ above which CSB occurs for 
fundamental-representation fermions (quarks) as found from a gap equation
that does {\em not} contain confinement effects; and (3) $g_c^2(0;adj)/(4
\pi )$, the same coupling for (quenched) adjoint fermions.  Because these
last two couplings differ from the first only by Casimir factors, the
critical couplings for gap equations are inversely
proportional to $C_2$, the quadratic Casimir eigenvalue for the fermions in
the gap equation, and so for QCD, $g_c^2(0; fund)/(4\pi
)=(9/4)g_c^2(0;adj)/(4 \pi )$.  However, just knowing this
 is not enough to settle the issue of whether CSB can or cannot take place
through a standard gap equation for quarks or adjoint
fermions; we need not just the ratios but also the values of the critical
couplings to compare with estimates from other QCD models of the couplings. 

There are in principle three possibilities:
\begin{eqnarray}
\label{3poss}
g^2(0) & < & g_c^2(0;adj)\\ \nonumber
g_c^2(0;adj) & < & g^2(0) < g_c^2(0;fund)\\ \nonumber
g_c^2(0;fund) & < & g^2(0)
\end{eqnarray}
In the first case only quarks can show CSB, and  confinement is necessary
for quark CSB.  In the second case (which is favored both from the lattice
data and from the estimates I give here)  there is CSB for both adjoint and
fundamental fermions. Confinement is again necessary for quark CSB, whose
transition temperature must be rather near that of 
  the deconfinement transition; the gap equation is largely
irrelevant, althouogh it may account for some separation in these two
transition temperatures..  CSB for adjoint fermions comes solely from
non-confining
effects as summarized in a gap equation, and adjoint-fermion CSB may or may
not extend above the
deconfinement transition.  In the third case, there is CSB for both kinds of
fermions,
and this ought to persist even if confiners such as center vortices are
removed from the lattice simulations.  Since there is more than a factor of
two between adjoint and fundamental critical couplings, the inequality in
the second case is a fairly broad one and great accuracy is not needed to
single out this case.

Fermion gap equations have a long history, beginning with the JBW equation
\cite{jbw}, and nearly all previous work that
does not address confinement issues makes three approximations:  1)  Gluons
are massless; 2) Landau gauge is used; 3) because to one-loop order in this
gauge vertex corrections are not ultraviolet-divergent, vertex corrections
are ignored.  Non-perturbative phenomena of low-energy QCD require a more
careful treatment that I will sketch here.  In the first place, infrared
slavery implies dynamic gluon mass generation, which can be only studied
effectively in the gauge-invariant pinch technique \cite{corn82,abp}.  In
the second place, because
fermion mass generation is an infrared effect there
may be important low-energy fermion-gluon vertex corrections.  Finally, the
possibility of such corrections requires a
more careful study of gauge invariance of the gap equation.  To some extent
these issues have been addressed before \cite{bijan,handh,cornpap91}, but
not in a context particularly useful here.  In the present work I include a
dynamical gluon mass in the gluon propagator; give  a
sketch of the derivation of a gauge-invariant gap equation using the pinch
technique (which seems not to have been addressed in detail before); and
use  the gauge technique to infer approximate low-momentum vertex
corrections that satisfy the correct Ward identities of the pinch
technique.  The simplest application of these principles, all that I report
here, yields a gap equation much like the JBW \cite{jbw} equation except
that the gluon is massive.

The work reported here is still very much in progress, and needs
considerable sharpening.  However, I believe it is already at the point
where one can qualitatively see and explain the differences between CSB for
quarks and for adjoint fermions.

\section{The functional Schr\"odinger equation for QCD}

This work has already been published \cite{corn07} so I will be brief here.
The vacuum wave functional of QCD is a gauge-invariant functional that I
write in the form:
\begin{equation}
\label{fsefunct}
\psi \{A_i^a(\vec{x})\}=e^{-S_3\{A_i^a(\vec{x})\}}
\end{equation}
in which $2S_3$ is a real gauge-invariant $d=3$ effective action (a factor
of two because $|\psi |^2$ is the weight function for constructing
vacuum expectation values).  It is constructed to satisfy
\begin{equation}
\label{fse}
H\psi = E_{vac}\psi
\end{equation}
with Hamiltonian
\begin{equation}
\label{hamil}
H=\int\{-\frac{1}{2}g^2(\frac{\delta}{\delta A_i^a})^2+
\frac{1}{4g^2} (G_{ij}^a)^2 \}\equiv 
\int [\frac{1}{2}(\Pi_i^a)^2]+V.
\end{equation}
The functional $S_3$ has infinitely many terms:
\begin{equation}
\label{s3form}
g^2S_3=\frac{1}{2!}\int \int A_i^a\Omega_{ij}A_j^a+\frac{1}{3!}\int \int
\int
A_i^aA_j^bA_k^c
\Omega_{ijk}^{abc}+\dots
\end{equation}
Gauge invariance requires \cite{corn88} that $\Omega_{ij}$ be conserved, and
any two successive terms in the expansion are related by ghost-free Ward
identities.
Even the simplest of these terms for a free gauge theory is not familiar as
an effective action, because it has a square root:
\begin{equation}
\label{frees3}
S_{3free}=\frac{1}{2g^2}\int A_i^a\sqrt{-\nabla^2}P_{ij}A_j^a.
\end{equation}
Here $P_{ij}$ is the usual transverse projector:
\begin{equation}
\label{pijdef}
P_{ij}=\delta_{ij}-\frac{\partial_i\partial_j}{\nabla^2}.
\end{equation}

My approach to the FSE (and later to the fermion gap equations) begins with
the fact that QCD, because of infrared slavery, undergoes dynamical gluon
mass generation \cite{corn76,corn81,corn82,cornpap,cornyan,abp,abp1}.  The
FSE
must display this fact; how does it do so?  It is easy to describe in a toy
model, an Abelian gauge theory with gauge-invariant mass $M$ put in by
hand.  The Hamiltonian is:
\begin{equation}
\label{abelham}
H_{Abel}=\int\{-\frac{1}{2}g^2(\frac{\delta}{\delta A_i})^2+
\frac{1}{4g^2} [(F_{ij})^2+2 m^2A_iP_{ij}A_j]\}
\end{equation}
The corresponding $S_3$ that exactly satisfies this Hamiltonian is:
\begin{equation}
\label{s3abel}
S_{3Abel}=\frac{1}{2g^2}\int A_i\sqrt{m^2-\nabla^2}P_{ij}A_j.
\end{equation}

Once I add the mass, a nonlocality appears from the transverse projector. 
But this is easily remedied, by introducing a scalar field in the mass
term:\footnote{Or one can begin with a simple gauge-dependent mass term
$\int mA^2$ and project out its gauge-independent part by integrating over
all gauge transformations of the gauge potential.} 
\begin{equation}
\label{localize}
S_{3mAbel}  =  \frac{m}{2g^2}\int [A_i-\partial_i\phi]^2
\end{equation}
and functionally integrating over not only the gauge potentials but also
over $\phi$ when constructing vacuum expectation values.  This is entirely
equivalent to using the non-local transverse projector.  Note that the mass
term by itself satisfies a FSE with the $F_{ij}^2$ term missing from the
Hamiltonian of Eq.~(\ref{abelham}).

Since we are interested in infrared effects, it is reasonable to make an
expansion in inverse powers of the mass $m$ (or equivalently in powers of
the gradient operator).  The leading term is $\mathcal{O}(m)$ and is just
the mass term $S_{3mAbel}$ itself.  However, a naive expansion runs into a
little bit of trouble, as we see by expanding the square root in the exact
Abelian solution:
\begin{equation}
\label{sqroot}
S_{3Abel}= \rightarrow \frac{1}{2g^2}\int A_i[m-\frac{\nabla^2}{2m}+\dots]
P_{ij}A_j.
\end{equation}
(Observe that the second and succeeding terms in the expansion are local; in
fact, the second term is, up to a factor $1/m$, the usual $F_{ij}^2$ term.)
If only these two terms are kept, the field has mass $\sqrt{2}m$ and not
$m$.  Higher-order terms not written must correct for this discrepancy.
Of course, no such expansion in $1/m$ is necessary for the Abelian case, but
it is for the non-Abelian case, and requires \cite{corn07} a somewhat
better approximation to the square root which is usable for momenta whose
components are comparable in size to $m$.

How does this go for the non-Abelian case?  I have argued for decades
\cite{corn74} that
one describes locally gauge-invariant gauge-boson mass generation through a
gauged non-linear sigma model (GNSM), analogous to the local mass action
$S_{3M}$ above.  To
simplify the notation I use the anti-Hermitean matrix gauge potential
\begin{equation}
\label{antiherm}
A_i=\frac{1}{2i}\lambda_aA_i^a,
\end{equation}
where the $\lambda_a$ are the standard Gell-Mann matrices, 
and covariant derivative
\begin{equation}
\label{covder}
D_i=\partial_i+A_i.
\end{equation}
Introduce a unitary matrix $U$, with the gauge transformation properties
$U\rightarrow VU$ when the gauge potential is transformed by:
\begin{equation}
\label{gaugetrans}
A_i\rightarrow VA_iV^{-1}+V\partial_iV^{-1}.
\end{equation}
Then the locally gauge-invariant GNSM mass term is\footnote{Since I am only
interested in infrared
effects I intepret the coupling as being evaluated at zero momentum.}:
\begin{equation}
\label{nabelmass}
S_{3m}=\frac{-m}{g^2}\int Tr[U^{-1}D_iU]^2.
\end{equation}
The non-covariant derivative $U\partial_iU^{-1}$ is the non-Abelian
generalization of the Abelian scalar $\partial_i\phi$; in fact, the GNSM
action can be written as:
\begin{equation}
\label{ahat}
S_{3m}=\frac{-m}{g^2}\int Tr[A_i-U\partial_iU^{-1}]^2.
\end{equation}  
One can, just as in the Abelian case, eliminate $U$ through its equations
of motion (that is, minimize $S_{3m}$), which are (after some non-trivial
algebra):
\begin{equation}
\label{ueqns}
[D_i, A_i-U\partial_iU^{-1}]=0.
\end{equation}
The perturbative solution has infinitely many terms, of which a few are
\cite{corn74}:
\begin{equation}
\label{pertsol}
 U=e^{\omega};\;\;\omega = \frac{-1}{\nabla^2}\partial \cdot A
+\frac{1}{\nabla^2}\left \{[A_i,\partial
_i\frac{1}{\nabla^2}\partial \cdot A]+\frac{1}{2}[\partial \cdot
A,\frac{1}
{\nabla^2}\partial \cdot A]+\cdots \right \}
\end{equation}
The linear term simply generates the transverse projector I have already
used in the Abelian case.  In addition, there are non-perturbative
solutions relevant for center vortices.

I claim that this GNSM mass term is the leading term in the $1/m$ expansion
of an effective action that capture the leading non-perturbative effect of
infrared slavery, which is dynamic gluon mass generation.  (It also captures
the structure of massless poles in the pinch-technique Schwinger-Dyson
equation yielding the mass dynamically.)  It is a good
candidate for the leading term of the effective action $S_3$ because it is
gauge-invariant; indeed, just as in the Abelian case, it comes from
projecting out the non-Abelian gauge invariant from the simple $A^2$ mass
term.   It is almost evident without calculation\footnote{Ref.
\cite{corn07} contains details about the calculation, using the pinch
technique and the gauge technique.} that the
next-leading term should be the usual Yang-Mills term, which is just the
non-Abelian gauge completion of the Abelian term shown in
Eq.~(\ref{sqroot}): 
\begin{equation}
\label{nextlead}
S_{3} =\frac{-m}{g^2}\int Tr[A_i-U\partial_iU^{-1}]^2-\frac{1}{4g^2m}
\int Tr G_{ij}^2+\dots
\end{equation} 
where $G_{ij}$ is the usual Yang-Mills field strength.   

The normalization of the second term follows from the fact that the
quadratic term in $S_3$ is just the Abelian action of Eq.~(\ref{sqroot}),
one copy for each gauge boson. But just as in the Abelian case this wrongly
yields a free-field mass of $\sqrt{2}m$ instead of $m$.  I have proposed
\cite{corn07} to cure this approximately by choosing a renormalization
factor $Z$ that best approximates the actual square root operator of the
FSE with a two-term expansion; the result for the approximate two-term
action $I_{d=3}=2S_3$ is:
\begin{equation}
\label{finalconj}
-2S_3\equiv -I_{d=3}=  \frac{2mZ}{g^2}\int
d^3xTr[U^{-1}D_iU]^2\}+\frac{Z}{mg^2}
\int d^3x TrG_{ij}^2+\mathcal{O}(m^{-3})
\end{equation}
with $Z\simeq 1.1-1.2$.

The final step is to compare this to the standard $d=3$ form of
gauge-invariant massive QCD for the given mass $m$, which is:
\begin{equation}
\label{cand3}  I_{d=3}=-\int
d^3x\left
\{\frac{1}{2g_3^2}TrG_{ij}^2+\frac{m^2}{g_3^2}Tr[U^{-1}D_iU]^2\right \}.
\end{equation}
Here the $d=3$ coupling $g_3^2$ has the dimensions of mass.  Comparing the
two forms of $I_{d=3}$ yields:
\begin{equation}
\label{compare2}
g^2=\frac{2Zg_3^2}{m}.
\end{equation}
Several
authors \cite{chk,cornyan,alexn,buchphil,corn1998,eber,karsch,hkr,ckp,naka}
have given either lattice or theoretical estimates of the dimensionless
ratio $m/Ng_3^2$ for gauge group $SU(N)$
with $N=2,3$.   The results are quite consistent with an  average value of  
$g_3^2/m\simeq 6.3/N$.  One can roughly convert the no-quark coupling in
Eq.~(\ref{compare2}) to three
light
flavors by multiplying the right side of Eq.~(\ref{compare2}) by 11/9, and
the resulting value of the strong coupling at zero momentum is:
\begin{equation}
\label{fseg2}
\frac{g^2(0)}{4\pi} \simeq 0.5.
\end{equation}

This estimate and other quite similar pinch-technique estimates
\cite{abp,abp1}, combined with phenomenology \cite{natale} that gives
somewhat higher values, suggests that
$\alpha_s(0)\simeq 0.6\pm 0.2$.  I will now see how this range of values
fits into fermion gap equations for CSB.

\section{Confinement, soliton condensates, and gap equations}

The first step is to understand the difference between gap equations that
purport to show the effects of confinement and those that do not.  I will
not do that in any detail here, but simply draw a few conclusions from the
fact that confinement comes from the long-range pure-gauge parts of center
vortices, which are quantum solitons of an effective action of the type
given in Eq.~(\ref{cand3}) or its $d=4$ extension.  Because of
complications having to do with integrating over center-vortex collective
coordinates, it is easiest to present the argument in $d=2$, where I will
use not the familiar action of $d=2$ Yang-Mills theory but rather the
effective action, with a mass term, of Eq.~(\ref{cand3}) in two dimensions
\cite{corn98}.  (I have argued \cite{corn07} that this is (possibly up to
an overall factor) the correct action for the $d=2+1$ FSE vacuum wave
functional, and that leaving out the mass term cannot be right; the reason
is that without the mass term Wilson loops of all representations show an
area law, while with it only $N$-ality $\neq$ 0 representations are
confined, which is correct for $d=2+1$.  But all this  is irrelevant to the
present argument.)  

The simplest center vortex is a soliton solution to the equations of motion
of the effective $d=2$ action:
\begin{equation}
\label{dis2}
A_j(x-a;K)=(2\pi
Q_K/i)\epsilon_{jk}\partial_k\{\Delta_m(x-a)-\Delta_0(x-a)\}.
\end{equation}
Here $Q_K,K=1\dots N-1$ is a generator of the Cartan subalgebra of $SU(N)$,
normalized so that $\exp  [2\pi iQ_K] $ is an element of the center, and
$\Delta_{m,0}$ are free propagators of mass $m,0$. The vector $a$ is a
collective coordinate for translations, and  I do not indicate collective
coordinates for group rotations.  A gluon propagator can be defined by
integrating the product of two soliton potentials over their common
collective coordinates:
\begin{equation}
\label{solprop}
\langle A^a_i(x)A^b_j(y) \rangle = \frac{\delta_{ab}}{N^2-1}\sum_a
\sum_{K=1}^{[N/2]}(-2)Tr(Q_J^2)A_i(x-a;K)A_j(y-a;K). 
\end{equation}

This propagator has a long-range part coming from the $\Delta_0$ term, and
the remainder is short-range.  The full  gluon propagator, with both terms,
can be used in the gap equation for the pinch-technique fermion propagator.
This gap equation is derived from an S-matrix element, or equivalently from
some complicated functional of a Wilson loop.  The long-range pure-gauge
parts are detected by their linkage with the Wilson loop.  If a vortex is
inside the loop it gives a non-trivial center element; otherwise it gives
unity.  Since there are no non-trivial elements for $N$-ality zero
representations, such as the adjoint, the adjoint fermion is completely
blind to them and
sees only the short-range parts.  So for quarks the gap equation should be
described with a gluon propagator containing the long-range pure-gauge
terms, which gives the long-range gluon propagator:
\begin{equation}
\label{shortrange}
\langle A^a_i(x)A^b_j(y) \rangle|_{long} = const.\delta_{ab}
(\Delta_0)_{ij}(x-y)
\end{equation}
where
\begin{equation}
\label{2dprop}
(\Delta_0)_{ij}(x-y)=\frac{1}{(2\pi )^2}\int d^2k(\delta_{ij}-k_ik_j/k^2)
\frac{e^{ik\cdot x}}{k^2}
\end{equation}
is the gauge propagator of $d=2$ QCD.  This propagator, singular at large
distances, not only confines quarks, it breaks CSB.  (Generally in $d$
dimensions the propagator behaves like $k^{-d}$, which I used in $d=4$  in
an earlier discussion \cite{corn80} of confinement and CSB.)

The remaining short-range part couples to all fermions with strength
proportional to the quadratic Casimir $C_2$, and has range $1/m$.  In fact,
these short-range soliton parts must sum to a standard massive gauge
propagator of the form (omitting the group labels):
\begin{equation}
\label{fullprop}
\Delta_{ij}(k) = P_{ij}(k)\frac{1}{k^2+m^2}+\frac{\xi k_ik_j}{k^4}.
\end{equation}
I have written this propagator in the form given by the pinch technique,
where the physical part is both gauge-invariant (unlike conventional
propagators) and transverse. The last term on the right is a necessary but
inert term depending on the chosen gauge that cannot enter any
pinch-technique physical prediction.  In particular, it must cancel out in
pinch-technique fermion gap equations.

\section{Fermion gap equations without confinement}

In this section I briefly mention the older gap equations, which are
oriented toward ultraviolet behavior.  Then I go onto newer equations that
treat the infrared regime of QCD more accurately, including a quick
discussion of a pinch-technique gap equation.  More details on these newer
gap equations will be published elsewhere.

The history of gap equations, from the Johnson-Baker-Willey (JBW) equation
of the sixties \cite{jbw} to work of the nineties, can be traced from
various specializations of an approximate gap equation for the CSB-breaking
running fermion mass $M(p^2)$:  
\begin{equation}
\label{jbw}
 M(p^2)=3C_2\int \frac{d^4k}{(2\pi
)^4}\frac{g^2(k^2)M(k^2)}{(k^2+M(k^2)^2)((p-k)^2
+m^2)}.
\end{equation}
An often-studied variant, and the only one I will consider explicitly, drops
the non-linear fermion mass terms in the denominator of the fermion
propagator on the right side of the equation, replacing $k^2+M(k^2)^2 $ by
$k^2$.  Trouble arises with any linearized gap equation that  has a
massless gluon propagator, because it is impossible to have a finite
fermion mass $M(0)$ at zero momentum.  Removing this problem by keeping the
full non-linear equation with a massless gluon and a fermion mass in the
denominator is not really a
good solution, since the only mass scale would have to come from the running
charge, which is essentially constant in the low-momentum regime where the
non-linear fermion mass is needed as an infrared cutoff.  In fact the
fundamental infrared cutoff is the gluon mass $m$ and I expect that
$M(0)\sim m$.

JBW used the linearized equation for QED, where $C_2\equiv 1$, the charge
does not run, the photon mass $m$ vanishes, and $M$ is the electron mass,
supposed to be generated spontaneously in massless QED.  The idea was to
show that this equation leads to a running mass vanishing at large
momentum, hence requiring no bare mass counterterm.
When this equation is used for QCD, $C_2$ is the quadratic Casimir
eigenvalue, $m$ is the gluon mass, and $g^2(k^2)$ is the running QCD
charge.  (In principle the dynamical gluon mass $m^2$ must run too
\cite{corn82}, but since the most important effect of this mass is in the
infrared region I will not include such running.)  

The master gap equation is a  
simplified form of the Schwinger-Dyson equations for the fermion
propagator $S(p)$, which has the form:
\begin{equation}
\label{fprop}
S^{-1}(p)=p\!\!/A(p^2)[1+iM(p^2)].
\end{equation}
As defined, the fermion ``mass" is essentially gauge-invariant, at least in
the sense that its ultraviolet anomalous dimension is gauge-invariant.  
  In practice, most workers specialize to the Landau gauge  
because large-momentum radiative corrections to the fermion-gluon vertex
are absent in one-loop order, so it is argued that in this gauge it should
be
a decent approximation to ignore vertex corrections and set (using the QED
Ward identity) $A(p^2)=1$ for all momenta.\footnote{In QED, with its small
coupling, ignoring higher-order
effects could well be justified.   In QCD, with its strong coupling, the
justification is that we are looking for infrared-dominated effects, so
large-momentum contributions should not be important.}  I will also use
Landau gauge and assume $A(p^2)$ does not change much with momentum,
although this requires a few words of justification for the pinch
technique/gauge technique approach given later.

The linearized fermion gap equation has the generic matrix form
\begin{equation}
\label{gengap}
M=g^2KM 
\end{equation}
where  
$K$ is the kernel, derived from the single skeleton graph for the inverse
fermion propagator.  There is always a chiral-symmetry-preserving solution
$M\equiv 0$, but we seek CSB-breaking solutions with $M\neq 0$.  If the
kernel is a well-behaved (finite-dimensional, bounded) matrix it is clear
that CSB can only occur if $g^2$ is sufficiently large; otherwise the
determinant $\det (1-g^2K)$ will not vanish.  Actually, $K$ (from, for
example, Eq.~(\ref{jbw})) is not that
well-behaved, but in the equations we use there is a critical coupling
$g_c^2$ marking the boundary between CSB and chiral symmetry preservation;
most students of gap equations give rather similar values for this critical
coupling.

There are two forms of criticality:  The first is based on a differential
equation derived from the gap equation.  Its solution may be well-behaved
for sufficiently small coupling, and then show unphysical features, such as
alternation of signs of the running fermion mass, for larger coupling.  The
second is based on the original linearized integral equation, which imposes
a boundary condition equivalent to consistency between the left-hand and
right-hand sides of the gap equation evaluated at zero momentum. 
Consistency fails if the coupling is too small.  Whether either or both 
criticality criteria hold depends on the gap equation, as I will show by an
explicit example.

\subsection{Ultraviolet behavior:  The JBW equation and variants}

Now I give some simple special cases, the first of which is well-known, of
the master gap equation of Eq.~(\ref{jbw}).  The first is the original JBW
equation used to study possible dynamical generation of the electron mass in
QED. The JBW equation sets $g^2$ to a constant and has no mass terms in the
propagators on the right side.
With the aid of
\begin{equation}
\label{momprop}
\bo \frac{1}{(p-k)^2}=-4\pi^2\delta (p-k)
\end{equation}
the JBW equation (now for QCD, so $C_2$ is reinstated) becomes the
differential equation
\begin{equation}
\label{jbw2}
M''+\frac{2M'}{p^2}+\frac{\lambda M}{p^4}=0
\end{equation}
where
\begin{equation}
\label{lambda}
\lambda = \frac{3C_2g^2}{16\pi^2}
\end{equation}
and the primes indicate derivatives with respect to $p^2$.  There are two
linearly-independent solutions:
\begin{equation}
\label{jbwsol}
M_{\pm}(p^2)=const.(p^2)^{\nu_{\pm}},\;\;\nu_{\pm} =\frac{1}{2}\{-1\pm
[1-4\lambda ]^{1/2}\}
\end{equation}
For small coupling the $\nu_-$ solution decreases roughly at large momentum
like $1/p^2$ and is commonly called
infrared-dominated; the $\nu_+$ solution corresponds to the original JBW
solution, which falls off very slowly at large momentum and is called
ultraviolet-dominated.  Although we may term one of the solutions
infrared-dominated, this does not mean that it can be used for small
momenta; in fact, both $M_{\pm}(0)$ diverge.  The infrared-dominated
solution is only useful in the ultraviolet.

These solutions $M_{\pm}$ are not appropriate for finding the ultraviolet
behavior in QCD, where the charge runs  at large momentum. 
This issue was first clarified by Lane \cite{lane}, who showed that
the master gap equation with a running charge and  no masses exactly
captures the ultraviolet behavior of the
fermion running mass.   At large momentum the gluon mass $m$ can be dropped
in the
denominators, and the running coupling has its usual behavior (at leading
order) 
\begin{equation}
\label{rung}
g^2(k^2)\simeq \frac{1}{b\ln (k^2/\Lambda^2)}
\end{equation}
where $b$ is the leading term in the $\beta$-function ($\beta
=-bg^3+\dots$) and  $\Lambda$ is the QCD mass.
   With the aid of
the renormalization group Lane showed that the ultraviolet behavior
appropriate to CSB is:
\begin{equation}
\label{lane}
M(p^2)\sim \frac{(\ln p^2)^a}{p^2};\;\;a=\frac{3C_2}{16\pi^2b}.
\end{equation}
I will recover this behavior later in a JBW-like equation with both a gluon
mass and a running charge.

\subsection{Critical couplings}

The simple JBW equation, with a non-running charge evaluated at zero
momentum, has a critical coupling at $\lambda = 1/4$, corresponding to a
critical coupling $g_c^2$ of value 
\begin{equation}
\label{critg}
 \frac{g_c^2}{4\pi}= \frac{\pi}{3C_2}.
\end{equation}
For couplings larger than critical the exponents $\nu_{\pm}$ become complex,
with $\nu_+^*=\nu_-$, and the asymptotic solutions both decay and
oscillate, for example, like   $\sim p^{-1}\cos  [\ln p]$.  There is
certainly no reason to accept as physical a running fermion mass that
alternates in sign. I have already noted that for a well-behaved kernel
$g^2$ must exceed a certain value for CSB to take place, yet for the simple
JBW equation criticality marks the onset of  apparently unphysical
behavior. So is criticality in the JBW equation at all related to CSB?  The
general answer is yes, although CSB may require a somewhat different value
from $g_c^2$.   
 
There are in fact a few reasons to believe  that $g_c^2$ above is indeed
close to the
true critical coupling for CSB.  For example, Miransky  \cite{miransky} and
his
collaborators \cite{fomin} have studied the positronium Bethe-Salpeter
equation  and find that if $g^2/(4\pi )\geq \pi /4$ tachyonic levels
appear.  They relate these to vacuum rearrangment and scale-breaking
phenomena, of which CSB is an example, and compare the 
  oscillatory behavior of equations like the JBW equation with supercritical
coupling to the quantum-mechanical problem of fall into the center or to 
the behavior of solutions to the 
Dirac equation with supercritical QED charge $Z\alpha >1$. In fact, the JBW
differential 
equation is nothing but the radial Schr\"odinger equation at zero angular
momentum and energy for an attractive potential $V(p^2\equiv r)$:
\begin{equation}
\label{fall}
V(r)=\frac{-\lambda}{r^2}.
\end{equation}
This potential shows fall into the center if $\lambda
\geq 1/4$, which is just the critical coupling given above.
 The QCD critical coupling of Refs. \cite{miransky,fomin} would be $\pi
/(4C_2)$, 
not much different from the JBW critical value.  Others (see \cite{higashi}
and references therein) claim that the JBW value is the critical coupling
for the pion Bethe-Salpeter equation to admit a massless pion.  So I will
assume that a critical coupling deduced from the gap equation is close to,
if somewhat below, the
critical coupling above which there is CSB.

\section{Infrared gap equations, gauge invariance, and the pinch technique}

The renormalization group cannot say anything about the behavior of $M(p^2)$
at low momentum, where (among other things) it becomes necessary to include
the effects of the gluon mass $m$ not only on the propagator but also on
the running coupling.   I and
others (see \cite{brodsky2}, which has many references to other works) claim
that in QCD there is a quasi-conformal infrared regime
where the running charge $g^2(k^2)$ is only slowly changing with momentum. 
Long ago  it was argued \cite{corn82} that  a decent approximation to the
running charge at both low (Euclidean) and  high momentum, with the right
sort of two-gluon threshhold,   is:
\begin{equation}
\label{corn82}
g^2(k^2)\simeq \frac{1}{b\ln (\frac{k^2+4m^2}{\Lambda^2})}.
\end{equation}
This quasi-conformal coupling runs very slowly at $k\ll m$.
Because  higher-order terms are neglected this expression cannot be more
than perhaps 10\% accurate. In the ultraviolet region a wide range of values
of $\Lambda$ does not change the coupling very much, but in the infrared
regime dimensional transmutation has taken place, with the zero-momentum
coupling determined by the ratio $2m/\Lambda$.  The reader can verify that
the choice
$m=0.5$GeV, $\Lambda=0.3$ GeV is within 10\% of a recent evaluation from
data \cite{davier} of the strong coupling at the masses of the $\tau$ and
$Z$, and (for three light flavors) gives $\alpha_s(0)\equiv g^2(0)/(4\pi
)\simeq 0.6$.   

Now consider the linear gap equation keeping  the gluon mass, but the
running charge is replaced by the fixed-point value $g^2(0)$.
There is no simple differential equation, but one can do the angular
integrals.  The resulting integral equation is: 
\begin{equation}
\label{jbwmass2}
M(p^2)=\int_0^{\infty}dk^2\frac{3g^2(0)C_2}{8\pi^2}M(k^2)K(p,k) 
\end{equation}
where
\begin{equation}
\label{kernel}
K(p,k)=K(k,p)=\frac{1}{p^2+k^2+m^2+[(p^2+k^2+m^2)^2-4p^2k^2]^{1/2}}.
\end{equation}

These equations can only be solved numerically, but they are closely related
to a solvable
 differential equation with a dominating kernel $\tilde{K}$, such that
$K\leq \tilde{K}$.  The new
kernel is:
\begin{equation}
\label{kerapprox}
K(p,k)\rightarrow \tilde{K}\equiv \frac{1}{2}[\frac{\theta
(p^2-k^2)}{p^2+m^2}
+\frac{\theta (k^2-p^2)}{k^2+m^2}].
\end{equation}
The new kernel $\tilde{K}$ is exactly equal to $K$ for $p>0,k=0$ (or
$k>0,p=0$), and $K,\tilde{K}$
are asymptotically the same at large momentum. When both momenta are
non-zero, $\tilde{K}$ is greater, by a maximum factor of about 1.3 at
$k^2=p^2=m^2$.  It should therefore be that the critical
value $g_c^2$ for $K$ is greater than that for $\tilde{K}$.   I expect a
further (modest) increase in the critical coupling because if I had
included a properly-running charge in the equation with kernel $K$ it would
further reduce this compared to the presently-considered kernel $\tilde{K}$
with a fixed charge.

The approximate version of the original (linearized) master gap equation,
using the kernel $\tilde{K}$ and fixed charge, is:
\begin{equation}
\label{jbwir}
M(p^2)=\frac{3\zeta_KC_2g^2(0)}{16\pi^2}\{\frac{1}{p^2+m^2}\int_0^{p^2}dk^2M(k^2)+   
\int_{p^2}^{\infty}\frac{dk^2}{k^2+m^2}M(k^2)\}
\end{equation}
where $\zeta_K$, a positive number between zero and one, measures the
discrepancy
between using the kernel $g^2(k^2)K$ and the kernel $g^2(0)\tilde{K}$.  One
should think of $\zeta_K$ as roughly measuring some sort of momentum average
of
the form
\begin{equation}
\label{zave}
\zeta_K\simeq \langle \frac{K(p,k)\ln (4m^2/\Lambda^2)}{\tilde{K}(p,k)\ln 
[(k^2+4m^2)/\Lambda^2]}\rangle.
\end{equation}
I estimate roughly that $\zeta_K\simeq 0.7-0.8$.

The differential equation emerging from Eq.~(\ref{jbwir}) is exactly the
same as Eq.~(\ref{jbw2}) with no gluon mass, except that the independent
variable is changed from $p^2$ to $p^2+m^2$ and $\lambda$ of
Eq.~(\ref{lambda}) is changed to $\zeta_K\lambda$:
\begin{equation}
\label{jbwmass}
M''+\frac{2M'}{p^2+m^2}+\frac{\zeta_K\lambda M}{(p^2+m^2)^2 }=0.
\end{equation}
  The solutions are:
\begin{equation}
\label{irjbw}
M_{\pm}(p^2)=const.(p^2+m^2)^{\nu_{\pm}},\;\;\nu_{\pm} =\frac{1}{2}\{-1\pm
[1-4\zeta_K\lambda ]^{1/2}\}
\end{equation}
and, unlike the massless-gluon solutions of Eq.~(\ref{jbwsol}), these are
finite at $p^2=0$.

Criticality occurs now at $\zeta_K\lambda = 1/4$, or
\begin{equation}
\label{zcrit}
\frac{g_c^2(0)}{4\pi}\simeq \frac{\pi}{3\zeta_KC_2}.
\end{equation}

As mentioned above, there is another criticality criterion.  I consider
evaluating the integral in Eq.~(\ref{jbwir}) at zero momentum:
\begin{equation}
\label{zeromom}
1=\frac{\zeta_K\lambda}{4}\int_0^{\infty}dk^2\tilde{K}(k,0)M(k^2)/M(0).
\end{equation}
This sets a boundary condition on the linear combination of solutions
$M_{\pm}$ of the differential equation, from which the solution of the
integral equation must be formed.
One might expect that if $\lambda$ is too small this criterion can never be
satisfied,
so the differential equation can be satisfied but not the corresponding
integral equation.  However, because both $M_+$ and $M_-$ are finite at the
origin and integrable over the kernel $\tilde{K}$ at infinity, it is in
fact always possible to find a linear combination that satisfies
Eq.~(\ref{zeromom}).  The problem is to find two   numbers $\alpha_{\pm}$
such that 
\begin{equation}
\label{alphabet}
\alpha_+M_+(0)+\alpha_-M_-(0)=1 
\end{equation}
and
\begin{equation}
\label{alphabint}
1=\frac{\zeta_K\lambda}{4}\int_0^{\infty}dk^2\tilde{K}(k,0)
[\alpha_+M_+(k^2)+\alpha_-M_-(k^2)].
\end{equation}
These 2$\times$2 linear equations are soluble for $\alpha_{\pm}$ except for
at most one value of $\zeta_K\lambda$.  (In the case at hand, $\alpha_{\pm}=
\nu_{\mp}/(\nu_{\mp}-\nu_{\pm})$; even the limit $\zeta_K\lambda = 1/4$
exists.) 
So the criterion of Eq.~(\ref{zeromom}) is not useful whenever the two
solutions to the differential equation are both finite at the origin and
integrable over the kernel at infinity.

The final change to be made in the gap equation is to replace $g^2(0)$ by
$g^2(k^2)$ in the $\tilde{K}$-equation given as Eq.~(\ref{jbwir}).  As long
as the momentum dependence of the running charge is on the integration
variable $k^2$ alone, a
differential equation can be found for any $k$-dependence.  This
differential equation is:
\begin{equation}
\label{jbwrun}
M''+\frac{2M'}{p^2+m^2}+\frac{3C_2\zeta'_K g^2(p^2)M}{16\pi^2(p^2+m^2)^2}=0
\end{equation}
Here $\zeta'_K$  accounts for the average difference between $K$ and
$\tilde{K}$ but not for the running  charge ({\em cf.} Eq.~(\ref{zave})). I
estimate $\zeta'_K\simeq 0.9$.   Using the running charge of
Eq.~(\ref{corn82}) gives the equation:
\begin{equation}
\label{jbwrun2}
M''+\frac{2M'}{p^2+m^2}+\frac{a\zeta'_K M}{(p^2+m^2)^2\ln [(p^2+4m^2)
/\Lambda^2]}=0
\end{equation}
where $a$ is Lane's constant, from Eq.~(\ref{lane}).

To my knowledge this is not a form of any standard differential equation,
but it reduces to one in two cases. For small momentum  if $p^2$ is dropped
compared to $4m^2$, the result is Eq.~(\ref{jbwmass}), already solved.  An
equation that is infrared-finite and asymptotically-exact for large
momentum results from   replacing $p^2+4m^2$ by $4(p^2+m^2)$, which yields 
a confluent hypergeometric equation.  

Unfortunately, this is not very accurate for small momentum, but since it is
asymptotically-exact for large momentum I will consider it briefly.    
Replace the running charge by:
\begin{equation}
\label{modcharge}
g^2(k^2)=\frac{1}{b\ln [4(k^2+m^2)/\Lambda^2]},
\end{equation}
As one should insist, $g^2(0)$ is unchanged by this modification, and the
large-momentum behavior is insensitive to $m$.

In addition to the correction factor  $\zeta'_K$, the last term of
Eq.~(\ref{jbwrun2}) should be multiplied by a factor $\zeta_g\geq 1$ that
attempts to correct for this mutilation of the running charge.  Think of
$\zeta_g$ as an average of the type:
\begin{equation}
\label{qave}
\zeta_g\simeq \langle \frac{\ln [4(p^2+m^2)/\Lambda^2]}{\ln
[(p^2+4m^2)/\Lambda^2
}].
\end{equation}

Now change to the new independent variable
\begin{equation}
\label{teq}
t=\ln [\frac{4(p^2+m^2)}{\Lambda^2}].
\end{equation}
With this factor, Eq.~(\ref{jbwrun2}) becomes:
\begin{equation}
\label{teqn}
\ddot{M}+\dot{M}+\frac{a\zeta_g\zeta'_K M}{t}=0
\end{equation}
where dots indicate $t$ derivatives. 
A simple estimate suggests that $\zeta_g\simeq 1.1-1.2$, so the product
$\zeta_g\zeta'_K$ is essentially unity within the accuracy to which I
aspire, and I drop this product.

 Equation (\ref{teq}) is a confluent hypergeometric equation (see, for
example, \cite{wwat}).  The two linearly-independent solutions
corresponding to Whittaker functions have different asymptotic behaviors. 
At large momentum the infrared-dominated solution goes like
\begin{equation}
\label{irdom}
M_{IR}(p^2)\rightarrow \frac{(\ln p^2)^a}{p^2}[1+\mathcal{O}(\frac{1}{\ln
p^2})]
\end{equation}
which is just the behavior found by Lane \cite{lane}.
The ultraviolet-dominated solution goes like
\begin{equation}
\label{uvdom}
M_{UV}\rightarrow (\ln p^2)^{-a}[1+\mathcal{O}(\frac{1}{\ln p^2})].
\end{equation}

Both of these solutions are finite at $p^2=0$ and both are integrable over
the kernel (including the running charge), so the same situation arises as
with the closely-related Eq.~(\ref{jbwir}):  Except for at most one value
of $a$, there is always a linear combination of Whittaker functions that
satisfies the zero-momentum consistency condition analogous to
Eq.~(\ref{zeromom}) that holds when the charge does not run.  

Our other criterion for criticality is the onset of zeros in $M(p^2)$.   The
confluent hypergeometric equation, Eq.~(\ref{teqn}),
shows critical behavior at $a=a_c=1$---not in the sense of singularities in
the solution, but, as for the original JBW equation, if $a>a_c$ zeros of
the mass function $M(p^2)$ set in \cite{tric}.  

The criticality condition $a_c=1$ is not really dynamical; it depends only
on the particle spectrum and gauge group (see Eq.~(\ref{lane})).  In QCD
with three light flavors I find for quarks that $a=4/9$, but for (quenched)
adjoint fermions, $a=1$.  It appears unlikely that the modified JBW
equation could lead to CSB for quarks, but it might well do so for adjoint
fermions, in view of all the approximations and uncertainties of our
development.  

I can, for what it is worth, convert the condition $a_c=1$  into a
criterion for the critical coupling, combining
Eqs.~(\ref{lane},\ref{corn82}) by eliminating the $\beta$-function
coefficient $b$.  The result is:
\begin{equation}
\label{gcrit2}
\frac{g_c^2(0)}{4\pi}=\frac{\pi}{3C_2}(\frac{4}{\ln (4m^2/\Lambda^2)}).
\end{equation}  
This estimate differs from that coming from the ultraviolet JBW equation by
the factor in parentheses.  This factor plausibly varies from 1 (at
$m\simeq 3.7 \Lambda$) to 2 (at $m\simeq 1.4 \Lambda$), given uncertainties
in both $m$ and the effective value of $\Lambda$ that works best for a
one-loop approximation.

\subsection{The pinch technique/gauge technique gap equation}

My final infrared gap equation is based on the gauge technique (Ref.
\cite{cornhou} and references therein), in which one ``solves" the Ward
identity for the fermion-gluon vertex in terms of the fermion propagator. 
The gauge technique is combined with the pinch technique,\footnote{See
\cite{papabin02} for a thorough discussion of the two-loop pinch-technique
fermion self-energy.} which is a way
(already mentioned) of finding off-shell Schwinger-Dyson equations, such as
the fermion gap equation, that are locally gauge-invariant.  These are not,
of course, the usual Schwinger-Dyson equations; pinch-technique proper
vertices and self-energies contain contributions from parts of graphs
naively unrelated to the proper vertex under study.  As a result, the Ward
identities of a non-Abelian gauge theory are modified; they are just the
naive Ward identities of a QED-like theory. 

\begin{figure}
\begin{center}
\includegraphics[height=4in]{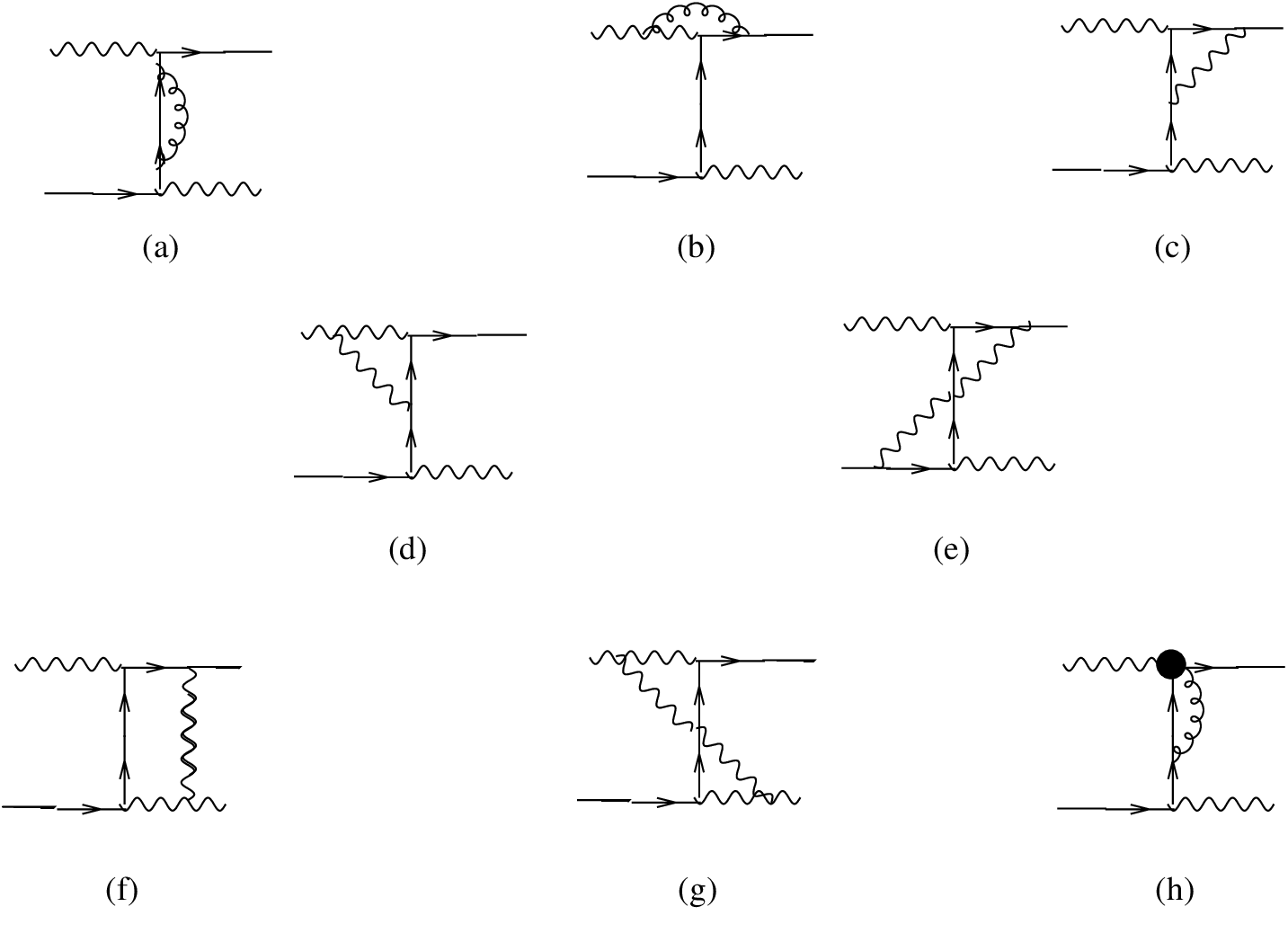}
\caption{\label{fig1-4} The on-shell quark-gluon amplitude for finding the
pinch-technique quark proper self-energy.  The last graph (h) is an example
of a new kind of graph emerging from pinching. }
\end{center}
\end{figure}

The pinch technique begins by setting up an on-shall S-matrix elements
containing the off-shell Green's function of interest---see
Fig.~\ref{fig1-4}.

Parts of numerators of some of these graphs contain longitudinal momenta
which, when they strike an elementary vertex, trigger an elementary Ward
identity of the form
\begin{equation}
\label{pinch}
k_{\mu}\gamma_{\mu}=S^{-1}(k)-S^{-1}(p-k).
\end{equation}
When an inverse propagator hits an extermal quark line it annihilates it,
and the inverse propagator with momentum of an adjoining internal quark
line simply removes that propagator, replacing it by unity.  The resulting
``pinch" yields graphs such as Fig.~\ref{fig1-4}(h).  A pinch can change
part of a vertex graph to a propagator graph, resulting in new terms that
must be added to the conventional fermion propagator.  Since the whole
S-matrix element is gauge-invariant, it turns out that the new fermion
propagator is gauge-invariant; gauge-dependent terms in the original
Feynman graphs are cancelled by these other pinch terms.  It is already
known \cite{corn81,corn82,binpap} that this procedure yields, in an
$R_{\xi}$ gauge,  a gauge-invariant gluon propagator of the type:
\begin{equation}
\label{gluonpropagator}
\hat{\Delta}_{\mu\nu}(k)=(\delta_{\mu\nu}-\frac{k_{\mu}k_{\nu}}{k^2})
\hat{d}(k)+\frac{\xi k_{\mu}k_{\nu}}{k^4}
\end{equation}
where $\hat{d}(k)$ is completely gauge-invariant (independent of $\xi$) and
the $\xi$ term receives no radiative corrections (other than wave-function
renormalization).  Not only do the longitudinal terms $\sim k_{\mu}$ cancel
out of
the S-matrix, they also cancel out of the fermion pinch-technique
propagator.  (Because of this cancellation proper vertices obey naive
ghost-free Ward identities.)     Although I have illustrated the pinch
technique only
to one loop in the figure, it is possible to extend it to all orders and to
non-perturbative phenomena (see \cite{papbin08} and references therein). 
For the gap equation the essential
non-perturbative phenomenon is a gluon mass, and I use
$\hat{d}(k)=1/(k^2+m^2)$.    

As is well-known \cite{binpap}, the pinch technique leads to Schwinger-Dyson
equations identical to those of the background-field method in the Feynman
gauge ($\xi =1$).  However, one can formulate the pinch technique in any
gauge \cite{pilaftsis} , in the sense that ghost-free Ward identities and
certain other structural elements of Schwinger-Dyson equations important
for the pinch technique in the Feynman gauge still hold.  I will, in
Pilaftsis' \cite{pilaftsis} sense of the pinch technique, use Landau gauge
as all other workers do, because ultraviolet corrections to the vertex are
unimportant to one loop in this gauge.  Because the Wasd identity relating
the vertex to the fermion propagator has no ghost terms, the same is true
(as in QED) for the coefficient $A(p^2)$  of $p\!\!/$ in the fermion
propagator [see Eq.~(\ref{fprop})].  In the infrared regime $A(p^2)$
should, like the running charge, not run much.  So I expect $A(p^2)\simeq
A(0)$ over a large momentum range, and will set it to unity.  (With the
gauge technique vertex, any constant $A(p^2)$ cancels out from the gap
equation.)

The gauge-technique solution for a pinch-technique form factor is an
infrared-valid
approximation to the form factor that is
asymptotically exact as the gluon momentum vanishes, and its ultraviolet
inaccuracies can be compensated for \cite{king}, although I will not
attempt that here.  Because its Ward identity is
ghost-free, just as in
QED,  all necessary formulas can be read off with straightforward
modifications from the QED work in \cite{king}.  One pinch technique/gauge
technique
improper form factor is (omitting irrelevant group factors and indices):
\begin{equation}
\label{imp}
F_{\mu}(p,p')=S(p)\Gamma_{\mu}(p,p')S(p') =\frac{1}{p^2-p'^2}
[(p\!\!/\gamma_{\mu}+\gamma_{\mu}p'\!\!\!\!/)S(p')-
S(p)(p\!\!/\gamma_{\mu}+\gamma_{\mu}p'\!\!\!\!/)]
\end{equation}
and obeys the QED-like Ward identity
\begin{equation}
\label{qedward}
(p-p')_{\mu}F_{\mu}(p,p')=S(p')-S(p).
\end{equation}
Eq.~(\ref{imp}) is not a unique choice for the form factor, but it has the
advantage of being an identity for a free massive theory, with
$\Gamma_{\mu}(p,p')=\gamma_{\mu}$ and $S^{-1}(p)=p\!\!/-iM$.  All other
choices for the pinch-technique form factor, as well as corrections needed
for  the ultraviolet behavior at loop level,  are identically conserved and
therefore differ from Eq.~(\ref{imp}) only by terms that vanish with at
least one more
power of $p-p'$ at small values of $p-p'$.

The Schwinger-Dyson (SD) equation is:
\begin{equation}
\label{sdgt}
S_0^{-1}(p)S(p)=1+\frac{g^2C_2}{(2\pi )^4}\int d^4k\gamma_{\nu}
\hat{\Delta}_{\nu\mu}(p-k)F_{\mu}(k,p)
\end{equation}
and with the gauge-technique form factor it is {\em linear} in the fermion
propagator.  In the pinch technique the gauge-boson propagator, which shows
dynamical mass generation, has the form given in
Eq.~(\ref{gluonpropagator}), except that the gauge-dependent term is
dropped.

To study CSB I take $S^{-1}(p)=p\!\!/\ $ and extract the coefficients of
$p\!\!/\ $ in the SD equation; this yields the gap equation.  Because the
form factor of Eq.~(\ref{imp}) is identically equal to that of a free
massive field theory and because the running fermion mass is not running
very fast in the infrared, the resulting pinch-technique equation at small
momentum (including small integration momentum $k$) is really the same as I
started with in Eq.~(\ref{jbw}), except that for small momenta I ignore the
running of masses and the coupling.  It is plausible that a nearly-correct
modification of the infrared pinch-technique equation that accounts for
ultraviolet effects is simply to let the masses and coupling run---cases I
have already reviewed or analyzed.  Actually, it is somewhat more
complicated than this, but the final analysis is much too elaborate to
discuss here.  

 The gauge technique incorporates infrared-important vertex corrections  and
the pinch technique yields a unique gauge-invariant gap equation in an
arbitrary gauge for the underlying Feynman graphs.  At the present rather
simple level of approximation to the pinch technique/gauge technique
gap equation,  the
deep-infrared gap equation, with plausible ultraviolet corrections, is the
same as found by many others over decades by ignoring vertex corrections
and working in the Landau gauge (of course, self-consistent at one-loop
level).

\section{Finite temperature CSB}

The full finite-temperature ($T$) gap equation, especially for the pinch
technique, has a number of complications that I have not dealt with as of
this writing, but
will save for a more detailed work.  These include separating the gluon
propagator into space-space, space-time, and time-time components, which
are no longer related by Lorentz invariance and which have two mass scales,
the magnetic and electric (Debye).  This complicates the application of
pinch-technique cancellation mechanisms \cite{chk}.   I will only present
here the crudest initial attempts at understanding finite-temperature CSB
with massive-gluon exchange, using what I call the
superconductor approximation, because it is in the same spirit as used in
the original BCS paper on superconductivity.  It amounts to saying
that the gluon propagator is relatively unchanged by temperature effects as 
long as the temperature is not too large compared to the phase-transition
temperature $T_c\simeq 170$ MeV.  The reason is that the gluon already has
a large $T=0$ mass of some 600 MeV, which is not changed drastically by
thermal effects at $T\sim T_c$.  However, the fermion mass steadily
decreases from its $T=0$ value, eventually vanishing at the CSB phase
transition\footnote{By mass I mean a quantity that violates chiral
symmetry, which the usual perturbative finite-temperature fermion ``mass"
does not.}.  The difference between fermionic and gluonic dynamical mass
generation at finite $T$ is that the effective coupling strength for
fermions is decreased as $T$ increases, by a factor something like $\tanh
(\beta \omega /2)$ where $\beta = 1/T$ and $\omega$ is a characteristic
energy scale; ultimately, the fermionic mass has to vanish as the coupling
diminishes. But the gluonic mass is increased by a factor roughly $\coth
(\beta \omega /2)$ (so the mass grows like $T$ at large $T$).  

In the superconductor approximation I convert the original zero-temperature
gap equation, Eq.~(\ref{jbw}), to finite temperature by the usual
replacement of the integral over the (Euclidean) time momentum $k_4$ by a
discrete frequency sum:
\begin{equation}
\label{p4int}
\int \frac{dk_4}{2\pi}\rightarrow \sum_{N=-\infty}^{\infty}\delta [k_4-
2\pi T(N+\frac{1}{2})].
\end{equation}
As mentioned above it is not quite correct simply to make this substitution
in  an equation such as (\ref{jbw}) where the vector-propagator kinematics
have been worked out at  zero temperature.  
A further approximation is to average certain momentum-dependent quantities
that vary fairly slowly by averages, which allows us to make contact with
an already-studied zero-$T$ equation, Eq.~(\ref{jbwir}).    The resulting
gap equation is then Eq.~(\ref{jbwir}) with the modified zero-momentum but
finite-$T$ coupling:
\begin{equation}
\label{newtcpl}
g^2(k=0,T=0)\rightarrow g^2(k=0,T)\langle \tanh (\frac{\beta
\omega_F}{2})\rangle
\equiv G^2(T)
\end{equation}
with the finite-$T$ coupling determined from the zero-momentum form of
Eq.~(\ref{corn82}) by using a plausible form for the temperature dependence
of the finite-$T$ gluon mass:
\begin{equation}
\label{gluont}
g^2(k=0,T)=\{b\ln [\frac{4m^2(T)}{\Lambda^2}]\}^{-1};\;\;m(T)=
m(T=0)\coth (\frac{\beta \omega_G}{2})
\end{equation}
where I choose for the average gluon frequency
\begin{equation}
\label{masst}
\omega_G=\frac{4\pi m(0)}{Ng^2(k=0,T=0)}
\end{equation} 
so as to give a simple but fairly accurate high-$T$ limit ({\em cf} the
value used in connection with the FSE of $m(T)=0.16Ng_3^2$, with
$g_3^2\simeq g^2(0)T$).  The fermion frequency $\omega_F$ is of the form:
\begin{equation}
\label{fermfreq}
\omega_F=\sqrt{\vec{k}^2+M^2(\vec{k})}.
\end{equation}

At the level of the simple approximations used  here, it is not possible to
predict with any accuracy the actual ratio of the temperature $T_{\chi A}$,
at which adjoint chiral symmetry breaking is restored, to the usual
deconfinement transition temperature $T_d$.  However, after some not very
interesting numerics involving plausible ranges of unknown and approximated
quantities, and using the superconductor approximation given here, it
appears possible that $T_{\chi A}>T_d$.  If this possibility survives more
detailed numerics then
adjoint chiral symmetry breaking survives deconfinement, as the lattice
simulations show.  

Of course, there is much more to learn about QCD in general from adjoint CSB
at finite temperature beyond our interests here, as the literature shows
\cite{perez, engels,basile}.  I hope to take up these questions in more
detail later.

 \section{Summary}

I use the FSE to estimate, using $d=3$ dynamical mass calculations, the
usual $d=4$ strong coupling at zero momentum:  $\alpha_s(0)\simeq 0.5$ (for
three light flavors), a value in accord with other estimates using the
pinch-technique Schwinger-Dyson equations.  This is   somewhat smaller
than phenomenological estimates of around 0.75.  I then consider
modern versions of the fermion gap equation both for quarks and for adjoint
fermions in QCD---made modern by adding dynamical gluon mass effects and by
consideration of the gauge-invariant pinch techniqe/gap technique gap
equation.  The final results, although not impressively accurate, suggest
that confinement is essential to break CSB for quarks, but that standard
fermion gap physics can explain CSB for adjoint fermions,  for some
(uncalculated) range of temperatures, possibly reaching above the
deconfinement temperature. 
These results are consistent with present-day lattice simulations.

\begin{flushleft}
{\Large \bf Acknowledgments}
\end{flushleft}

My attendance at this Symposium was partially supported by the UCLA Council
on Research.

\end{document}